\documentclass{ws-mplb}

\begin{document}

\markboth{Xue-Wen Wang et al.}{Modeling citation networks based on vigorousness and dormancy}

%
\catchline{}{}{}{}{}
%

\title{Modeling citation networks based on vigorousness and dormancy}

\author{Xue-Wen Wang, Li-Jie Zhang, Guo-Hong Yang}

\address{Department of Physics, Shanghai University, Shanghai 200444, China}

\author{Xin-Jian Xu}

\address{Department of Mathematics, Shanghai University, Shanghai 200444, China\\Institute of Systems Sciences, Shanghai University, Shanghai 200444, China\\
xinjxu@shu.edu.cn}

\maketitle

\begin{history}
\received{(Day Month Year)}
\revised{(Day Month Year)}
\end{history}

\begin{abstract}
  In citation networks, the activity of papers usually decreases with age and dormant papers may be discovered and become fashionable again. To model this phenomenon, a competition mechanism is suggested which incorporates two factors: vigorousness and dormancy. Based on this idea, a citation network model is proposed, in which a node has two discrete stage: vigorous and dormant. Vigorous nodes can be deactivated and dormant nodes may be activated and become vigorous. The evolution of the network couples addition of new nodes and state transitions of old ones. Both analytical calculation and numerical simulation show that the degree distribution of nodes in generated networks displays a good right-skewed behaviour. Particularly, scale-free networks are obtained as the deactivated vertex is target selected, and exponential networks are realized for the random-selected case. Moreover, the measurement of four real-world citation networks achieves a good agreement with the stochastic model.
\end{abstract}

\keywords{Complex networks; citation networks; degree distribution.}
\section{Introduction}

The citation patterns of scientific publications can be simplified into a citation network with nodes representing scientific articles published in journals and edges mimicking citations from one article to another published previously \cite{RS05}. Citation networks are valuable to uncover the dynamics of scientific publications and have been studied for a long time \cite{ER90}. A particularly noteworthy contribution was a study in $1965$ by de Solla Price \cite{deSP65}, who proposed the so-called \lq\lq cumulative advantage\rq\rq mechanism, that is, a paper which has been cited many times is more likely to be cited again than others which have been little cited. The cumulative advantage is based on the idea of \lq\lq rich get richer\rq\rq suggested by Yule \cite{YGU25} and Simon \cite{SHA55}. And the criterion now is widely known as the \lq\lq preferential attachment\rq\rq in the framework of currently fashionable evolving network models, proposed by Barab\'{a}si and Albert in $1999$ \cite{BA99}. By employing growth and preference, the Barab\'{a}si-Albert (BA) model provides a natural explanation for the scale-free behavior observed in many realistic systems. Recently, Clauset et al. \cite{CSN09} proposed a statistical framework for determining power-law tails of various data sets, in accordance with the conclusion of Redner \cite{RS98}.

In the study of citation networks, one of the most important topics is the characterization of the probability distribution of the number of citations received by a paper and the design of simple microscopic models to reproduce the real-world distribution \cite{RFC08}. Many empirical studies in citation networks have proved that age may be one of the most important mechanisms that determines the statistical properties of the growing network \cite{DM00,KE02,ZWZ03,VA03,HS06,TL06,WYY08,XZ09,YS09,XF11,GS12}. To investigate the effect of age on network evolution, the BA model has been modified by incorporating time dependence in citation networks. Dorogovtsev and Mendes \cite{DM00} studied the case that the probability of an old node attached by a newcomer is not only proportional to its degree $k$ but also to a power of its age $\tau^{-\alpha}$ (where $\tau$ is the age of a node). They found that the resulting network shows scale-free (SF) behavior only in the region $\alpha < 1$. For $\alpha > 1$, the degree distribution $P(k)$ is exponential. One the other hand, Klemm and Egu\'{i}luz \cite{KE02} proposed a degree-dependent deactivation network model which is highly clustered and retains the power-law distribution of the node's degree.

Most previous studies only consider the irreversible impact of age, such as gradual aging \cite{DM00} and absolute deactivation \cite{KE02}. In the real world, however, there is a universal phenomenon called ``delayed recognition'', that is, papers did not seem to achieve any sort of recognition until some years after their original publication \cite{VanR04,BQL05}. The question therefore arises as to whether such process can be explained or expected by the network theory. With the advance of the theory of complex networks, scientists can understand and describe real systems more subtly \cite{XXL06,WH09,WY09,WCHLH09,CRZH09,WYY11,ZLL12}.
In this paper we express the notion of the delayed recognition in terms of an evolving network model with transitions of nodes' states to answer this question. Intuitively, we suggest that the activity of a node is the result of the competition of two factors: vigorousness and dormancy. For vigorousness, supposing that a new published paper or an old paper, its ability of receiving citations from others increases gradually with time. Whereas for dormancy, it describes the deactivation of the paper and being slept. The evolution of the network couples addition of new nodes and state transitions of old ones. It is found that the degree distribution of the resulting network depends on the transition probability. Furthermore, we study four real-world citation data and notice the good agreement with present model.

\section{Model}

\begin{figure}
\resizebox{\columnwidth}{!}{\includegraphics{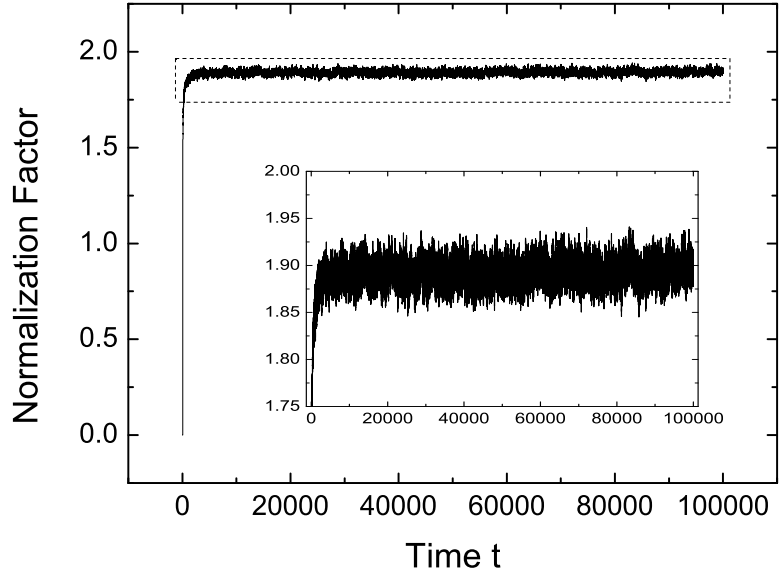}}
\caption{Illustration of the normalization factor $\gamma-1$ as a function of time $t$
with the parameters $m=20$ and $\alpha=m+2$. The amplified version can be seen in the
inset. The data points correspond to system size $N=10^5$, and each is obtained as an
average of 200 independent runs.} \label{fig1}
\end{figure}

The evolution process starts with an initial network of a small number $m_0$ of isolated nodes, in which $m$ ($m<m_0$) nodes are vigorous. Motivated by previous research \cite{KE02,XF11}, at each time step, the dynamics runs as follows.

(i) Adding a new node $i$ with $m$ outgoing links that are attached to previously existing $m$ vigorous nodes. We assume that $m$ is the average number of references per article. The in-degree of a node, i.e., the number of edges pointing to it, is denoted by $k^{'}$. At first, the in-degree of the newcomer is $k_{i}^{'}=0$. Each selected vigorous node $j$ receives exactly one incoming edge, thereby $k_{j}^{'} \rightarrow k_{j}^{'} +1$. Since the out-degree of each node is always $m$, the total degree of a node is $k=k^{'}+m$.

(ii) Activating the new node $i$, which means that the new published paper is always assumed to be vigorous at first.

(iii) Awakening one of the previously existing dormant nodes. For simplicity, we assume that each dormant node is chosen uniformly to be activated.

(iv) Deactivating two of the vigorous nodes. The probability of a vigorous node $j$ being deactivated is given by
\begin{equation}\label{probdeact}
\nu_{k'_j}=\frac{\gamma-1}{\alpha+k'_j},
\end{equation}
where $\alpha>0$ is a preferential factor reflecting the initial attractiveness of different fields, and the normalization factor is defined as $\gamma-1=[\sum_{l\in{\Lambda}}{1/(\alpha+k'_{l})}]^{-1}$. The summation runs over the set $\Lambda$ of the currently vigorous nodes. Eq.~(\ref{probdeact}) means that the most cited paper is less possibility to be forgotten.

According to the model definition, vigorous nodes may become dormant ones gradually, which can be explained as a collective ``forgetting''. At the same time, dormant nodes may be awaked and receive links from subsequent node again, which can be considered as the recognition of ``forgotten'' papers.

\section{Degree distribution}

Denoting $A_{k'}^{t}$ the number of vigorous nodes with in-degree $k'$ at time $t$, one can write out the differential equation
\begin{equation}\label{eqevolve}
\frac{\partial{A_{k'+1}^{t}}}{\partial{t}}=(1-2\nu_{k'}^{t})(A_{k'}^{t}+\mu_{k'}^{t})-A_{k'+1}^{t}=\left(1-2\frac{\gamma-1}{\alpha+k'}\right)(A_{k'}^{t}+\mu_{k'}^{t})-A_{k'+1}^{t}
\end{equation}
for network evolution, where $\mu_{k'}^{t}$ is the activation probability.
On the right-hand side of Eq.~(\ref{eqevolve}), the first term contributes to $A_{k'+1}^{t+1}$. This term accounts for two processes: the vigorous nodes with in-degree $k'$ at time $t$ is not deactivated and will be connected to the new node in the next time step, and an activation node with in-degree $k'$ at time $t$ is not deactivated and will be attached by the new node in the next time step.

We investigate the behavior of $\gamma-1$ in time evolution. Figure \ref{fig1} shows the relationship between the normalization factor $\gamma-1$ and time $t$. We find that $\gamma-1$ approaches a stable value with certain fluctuations as soon as the evolution of the network starts. It is assumed that the fluctuations
of the normalization factor $\gamma-1$ are small enough, such that $\gamma$ may be treated as a constant. Imposing the stationary condition $\partial{A_{k'}^{t}}/\partial{t}=0$, one obtains
\begin{equation}\label{eqrecur}
A_{k'+1}-A_{k'}=-2\frac{\gamma-1}{\alpha+k'}A_{k'}+\left(1-2\frac{\gamma-1}{\alpha+k'}\right)\mu_{k'}^{t}.
\end{equation}
The probability of a dormant node being activated is assumed to be uniform, so $\mu_{k'}^{t}$ takes the form
\begin{equation}\label{probact}
\mu_{k'}^{t}=\frac{N_{k'}^{t}}{m_0+t-m-1},
\end{equation}
where $N_{k'}^{t}$ represents the number of dormant nodes with in-degree $k'$ at time $t$. For large $t$, the total number of nodes in the network is approximately equal to the number of dormant nodes, and the overall in-degree distribution $n_{k'}$ can be approximated by considering the dormant nodes only. Thus, we obtain the relationship
\begin{equation}\label{relation1}
\mu_{k'}^{t}=n_{k'},
\end{equation}
and $n_{k'}$ can be calculated as the rate of the change of vigorous nodes $A_{k'}$,
\begin{equation}\label{relation2}
n_{k'}=A_{k'}-A_{k'+1}.
\end{equation}
Substituting Eqs.~(\ref{relation1}) and (\ref{relation2}) into Eq.~(\ref{eqrecur}), one obtains
\begin{equation}\label{expressak}
A_{k'}={A_{0}}\prod\limits_{i=0}^{k'-1}\frac{\alpha-2\gamma+2+i}{\alpha-\gamma+1+i}=
{A_{0}}\exp\left[\sum\limits_{i=0}^{k'-1}\ln\left(1+\frac{1-\gamma}{\alpha-\gamma+1+i}\right)\right],
\end{equation}
where $k'\geq1$ and the boundary value $A_0$ is equal to $1$ reflecting the constant addition of newcomers with initial $k'=0$. The analytical solutions corresponding to different $\alpha$ are given in the following.

\begin{figure}
  \includegraphics[width=\columnwidth]{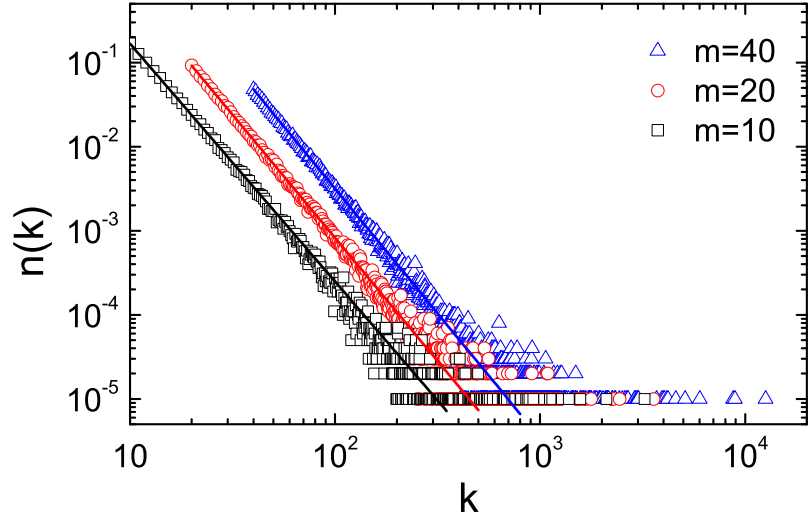}
  \caption{(Color online) Degree distributions of nodes of generated networks in case of $\alpha=m+2$ for $m=10$ (square), $20$ (circle) and $40$ (triangle), respectively. The size of networks is $N=10^5$. The solid lines are least-squares fits based on the form of Eq.~(\ref{distribpow}).}
  \label{fig2}
\end{figure}

(i) In the case of samll $\alpha$ and $\alpha\geq{m}$, Eq.~(\ref{expressak}) can be written as
\begin{equation}\label{expressak1}
A_{k'}\simeq(\alpha-\gamma+1)^{\gamma-1}(\alpha-\gamma+1+k')^{-(\gamma-1)},
\end{equation}
using the approximately logarithmic Taylor expansion, and then the overall in-degree distribution $n_{k'}$ is
\begin{equation}\label{distribpow}
n_{k'}=-\frac{dA_{k'}}{dk'}=c(\alpha-\gamma+1+k')^{-\gamma},
\end{equation}
where $c=(\gamma-1)(\alpha-\gamma+1)^{\gamma-1}$ is the normalized factor. The exponent $\gamma$ can be obtained from a self-consistency condition $m=\int_0^{\infty}k'n_{k'}dk'$, which gives
\begin{equation}
\gamma=1+\frac{m+\alpha}{m+1}.
\end{equation}
It can be seen that the exponent $\gamma$ depends on the parameters $\alpha$ and $m$. If $\alpha=m+2$, one has $\gamma=3$. Figure~\ref{fig2} shows the total degree distribution obtained by simulating the model for $10^5$ time steps. As expected, we obtain power-law distributions with best-fitted exponent $\gamma$ equal to $2.82(9)$, $2.90(9)$, and $2.96(5)$, corresponding to $m=10$, $20$, and $40$, respectively.

\begin{figure}
  \includegraphics[width=\columnwidth]{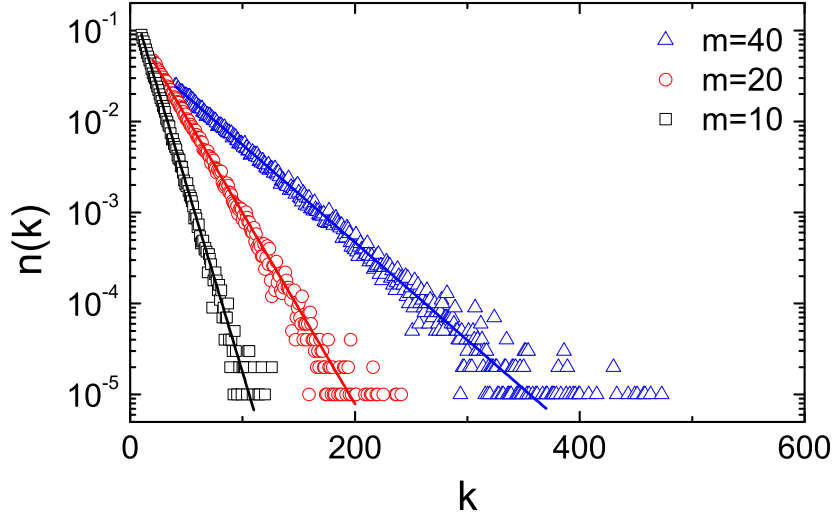}
  \caption{(Color online) Degree distributions of nodes of generated networks in case of $\alpha\rightarrow\infty$. The solid lines are least-squares fits based on the form of Eq.~(\ref{distribexp}).}
  \label{fig3}
\end{figure}

(ii) In the case of $\alpha\rightarrow\infty$, the deactivation probability $\nu_{k'}$ is independent of $k'$, which means that each of the $m+2$ vigorous nodes
will be deactivated with the same probability $1/(m+2)$. Thus, Eq.~(\ref{expressak}) can be written as
\begin{equation}
A_{k'}\simeq\exp\left[k'\ln\frac{\alpha-2\gamma+2}{\alpha-\gamma+1}\right]=\left(\frac{m}{m+1}\right)^{k'}.
\label{expressak2}
\end{equation}
Then, the overall in-degree distribution $n_{k'}$ is
\begin{equation}
n_{k'}=-\frac{dA_{k'}}{dk'}=\ln\left(\frac{m+1}{m}\right)\left(\frac{m}{m+1}\right)^{k'}.
\label{distribexp}
\end{equation}
To obtain the total degree distribution, we rewrite the above equation as
\begin{equation}
n_{k}=\ln\left(\frac{m+1}{m}\right)\left(\frac{m}{m+1}\right)^{k-m},
\end{equation}
where $k=k'+m$. Thus, the distribution is exponent decay. In Fig.~\ref{fig3}, we plot the total degree distribution of the simulated networks for $m=10$, $20$, and $40$, respectively. As expected, we obtain exponential distributions with best-fitted exponent $m/(m+1)$ being $0.90(9)$, $0.95(2)$, and $0.97(5)$,
corresponding to $m=10$, $20$, and $40$, respectively.

\begin{figure}
  \includegraphics[width=\columnwidth]{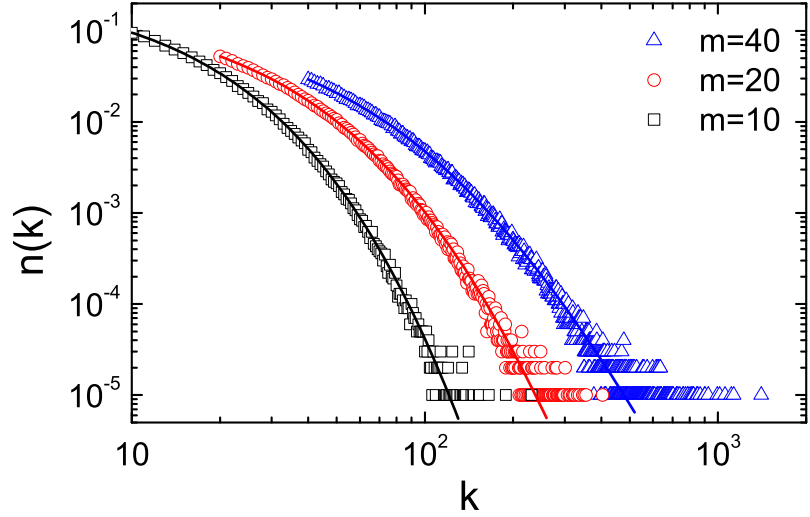}
  \caption{(Color online) Degree distributions of nodes of generated networks in case of $m\ll\alpha<\infty$. The solid lines are least-squares fits based on the form of Eq.~(\ref{distribgen}).}
  \label{fig4}
\end{figure}

(iii) In the case of $m\ll\alpha<\infty$, $A_{k'}$ can be represented by Eq.~(\ref{expressak2}) when $k'$ is small, and by the approximately logarithmic Taylor expansion when $k'$ is large.
Therefore, there exists a tipping point $k_c$ in the degree distribution. In the case of $k'$ being smaller than $k_c$, Eq.~(\ref{expressak}) can be simplified to
\begin{equation}
A_{k'}\simeq\left(\frac{m}{m+1}\right)^{k'}.
\end{equation}
While $k'$ is larger than $k_c$, Eq.~(\ref{expressak}) reduces to
\begin{equation}
A_{k'}\simeq\left(\frac{\alpha-2\gamma+2}{\alpha-\gamma+1}\right)^{k_c}(\alpha-\gamma+1+k_c)^{\gamma-1}\times(\alpha-\gamma+1+k')^{-(\gamma-1)}.
\end{equation}
Combining above two expressions, one can obtain the overall in-degree distribution $n_{k'}$
\begin{equation}\label{distribgen}
n_{k'}=(\gamma-1)\left(\frac{\alpha-2\gamma+2}{\alpha-\gamma+1}\right)^{k_c}\times(\alpha-\gamma+1+k_c)^{\gamma-1}(\alpha-\gamma+1+k')^{-\gamma}.
\end{equation}
In Fig.~\ref{fig4}, we plot the total degree distribution of the generated networks with parameters $\alpha=200$ for $m=10$, $20$, and $40$, respectively. All the plots are right-skewed, in agreement with the theoretical prediction.

\section{Comparison with empirical data}

\begin{figure}
  \includegraphics[width=\columnwidth]{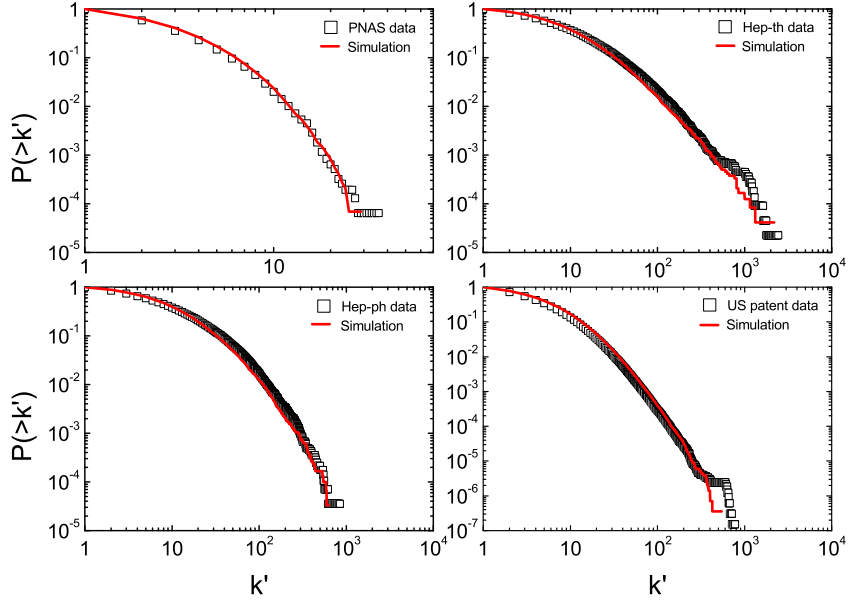}
  \caption{(Color online) Comparison of empirical networks with simulation results of the present network model. Parameters of simulations for different values $m_0=3145$, $m=2$, $\alpha=24$; $m_0=631$, $m=13$, $\alpha=13$; $m_0=2117$, $m=13$, $\alpha=18$; $m_0=470,978$, $m=5$, $\alpha=12.5$ correspond to PNAS, Hep-th, Hep-ph and U.S. Patent data, respectively.}
  \label{fig5}
\end{figure}

To examine present model, we utilize four empirical data from citation networks.

(i) PNAS data~\cite{RSC12}, which contains $23,572$ articles and $40,853$ edges published by the proceedings of the National Academy of Sciences (PNAS) of the United States of America from $1998$ to $2007$.

(ii) Hep-th data~\cite{GGK03}, which comes from preprints posted on arxiv.org, and covers papers in the period from January $1992$ to April $2003$ ($124$ months). It contains $27,770$ papers and $352,807$ edges.

(iii) Hep-ph data~\cite{GGK03}, which comes from preprints posted on arxiv.org, and covers papers in the period from January $1992$ to April $2003$ ($124$ months). It contains $34,546$ papers and $421,578$ edges.

(iv) U.S. Patent data~\cite{LKF05}, which is maintained by the National Bureau of Economic Research. The data includes all citations made by patents granted between $1975$ and $1999$, and contains $3,774,768$ nodes and $16,518,948$ edges.

Figure~\ref{fig5} shows the comparison of in-degree statistics of four citation networks with numerical results of generated networks. Our model has three parameters: the average out-degree $m$, the initial isolated nodes $m_0$ and the attractiveness bias $\alpha$. Since $N'$ and $E'$ are fixed, we assume that the value of $m$ is an integer and slightly larger than the average number of citations given out by all the papers in the empirical networks, and $m_0$ equals $N'-E'/m$. We scan the regions of $\alpha\in[0,30]$ with the increment $0.5$ and gain $\alpha$ by the best fit for the empirical data. The fits of PNAS, Hep-th and Hep-ph are averaged over $200$ independent realizations and $10$ independent runs for U. S. Patent~\cite{WH09,RSC12}. Although the empirical networks are different in nature, all the cumulative in-degree distributions follow a right-skewed decay which shifts from an exponential to a power law. Table~\ref{table11} shows empirical data on the citation distribution of papers and assessed parameters $m_0$, $m$ and $\alpha$ by simulation, and one notices the good agreement.

\begin{table}
  \caption{Basic statistics of PNAS, Hep-th, Hep-ph and U.S. Patent data. $N$, $E$ and $\overline{k}$ denote the number of nodes, edges and average out-degree in four empirical networks, respectively. $N'$, $E'$, $m_0$, $m$ and $\alpha$ are parameters for simulated networks. $N'$ and $E'$ denote the number of nodes and edges of the networks. $m_0$ represent the initial isolated nodes. $m$ and $\alpha$ represent the average out-degree and the constant bias in the networks.}
  \begin{center}
    \tabcolsep=9pt  
    \renewcommand\arraystretch{1.2}  
    \begin{tabular}{cllll}
        \hline
            $\frac{Measures}{networks}$ & PNAS & Hep-th & Hep-ph & U. S. Patent \\
        \hline
        $N$ & 23,572 & 27,770 & 34546 & 3,774,768 \\
        $E$ & 40853 & 352,807 & 421578 & 16,518,948 \\
        $\overline{k}$ & 1.7 & 12.7 & 12.2 & 4.4 \\
        \hline
            $N'$ & 23,572 & 27,770 & 34546 & 3,774,768 \\
            $E'$ & 40853 & 352,807 & 421578 & 16,518,948 \\
            $m_0$ & 3145 & 631 & 2117 & 470,978 \\
            $m$ & 2 & 13 & 13 & 5 \\
            $\alpha$ & 24 & 13 & 18 & 12.5 \\
        \hline
    \end{tabular}
  \end{center}
  \label{table11}
\end{table}

\section{Conclusion}

In summary, we have proposed a simple model for citation networks to explain the phenomenon of delayed recognition in the life of a article which usually begins lesser, rises to peak, and then diminishes. The growth dynamics of the network is governed by the state transition. We found that the average number of references per paper $m$ and the initial attractiveness of different fields $\alpha$ determine the topological structure of the generated network. If the value of $\alpha$ is selected appropriately as $m+2$, the deactivation probability $\nu_k$ is a linear preferential one, which leads to a power-law degree distribution with the exponent $\gamma=3$. Whereas for $\alpha$ tends to $\infty$, the vigorous nodes are selected to be deactivated with the uniform probability, and the model gives rise to an exponential degree distribution with the exponent only depending on $m$. Between the two regimes, the distribution gradually shifts from the exponential to the power law. To examine theoretical prediction, we compared the degree distribution with empirical citation networks and noticed a good agreement. So the present model
provides a new way to understand citation networks with age.

\section*{Acknowledgments}

This work was supported by the Innovation Program of Shanghai Municipal Education Commission (13YZ007) and the Specialized Research Fund for the Doctoral Program of Higher Education under No. 20093108110004.

\end{document}